\documentclass[sigconf,9pt]{acmart}
\usepackage{booktabs}
\usepackage{multirow}
\usepackage{hyperref}
\usepackage{microtype}
\usepackage{algorithm}
\usepackage{algorithmic}

\AtBeginDocument{%
  }

\copyrightyear{2025}
\acmYear{2025}
\setcopyright{rightsretained}
\acmConference[ASPDAC '25]{30th Asia and South Pacific Design Automation
Conference}{January 20--23, 2025}{Tokyo, Japan}
\acmBooktitle{30th Asia and South Pacific Design Automation Conference
(ASPDAC '25), January 20--23, 2025, Tokyo,
Japan}\acmDOI{10.1145/3658617.3697594}
\acmISBN{979-8-4007-0635-6/25/01}

\begin{document}


\title{DeepSeq2: Enhanced Sequential Circuit Learning with Disentangled Representations}

\author{Sadaf Khan}
\affiliation{
  \institution{The Chinese University of Hong Kong}
}

\author{Zhengyuan Shi}
\affiliation{
  \institution{The Chinese University of Hong Kong}
}

\author{Ziyang Zheng}
\affiliation{
  \institution{The Chinese University of Hong Kong}
}

\author{Min Li}
\affiliation{
  \institution{Huawei Noah's Ark Lab}
}

\author{Qiang Xu}
\affiliation{
  \institution{The Chinese University of Hong Kong}
}







\begin{abstract}
Circuit representation learning is increasingly pivotal in Electronic Design Automation (EDA), serving various downstream tasks with enhanced model efficiency and accuracy. One notable work, DeepSeq, has pioneered sequential circuit learning by encoding temporal correlations. However, it suffers from significant limitations including prolonged execution times and architectural inefficiencies. To address these issues, we introduce DeepSeq2, a novel framework that enhances the learning of sequential circuits, by innovatively mapping it into three distinct embedding spaces—structure, function, and sequential behavior—allowing for a more nuanced representation that captures the inherent complexities of circuit dynamics. By employing an efficient Directed Acyclic Graph Neural Network (DAG-GNN) that circumvents the recursive propagation used in DeepSeq, DeepSeq2 significantly reduces execution times and improves model scalability. Moreover, DeepSeq2 incorporates a unique supervision mechanism that captures transitioning behaviors within circuits more effectively. 
DeepSeq2 sets a new benchmark in sequential circuit representation learning, outperforming prior works in power estimation and reliability analysis.
\end{abstract}
\maketitle
\vspace{-4pt}\section{Introduction}
Circuit representation learning has recently gained significant traction within the Electronic Design Automation (EDA) field. This paradigm  involves two critical phases: pre-training a model on a comprehensive dataset to capture generic features and fine-tuning on a more focused dataset to refine these features for specific applications \cite{han2021pre,brown2020language}. 
This dual-phase approach ensures that models benefit from both a broad understanding gained during pre-training and the nuanced insights required for particular tasks.

To date, various netlist-level representation learning strategies have been proposed, predominantly focusing on either the structural or functional attributes of netlists \cite{he2021graph,li2022deepgate,wang2022functionality,shi2023deepgate2,wu2023gamora,khan2023deepseq}. Among these, \textit{DeepSeq} 
stands out as a pioneering framework for learning sequential circuits \cite{khan2023deepseq}. 
It employs a specialized graph neural network (GNN) with a custom propagation scheme to discern the temporal correlations between gates and memory elements (e.g., flip-flops (FFs)). It also supervise each component with logic-1 and transition probabilities, complemented by a novel dual attention aggregation function. 




Despite demonstrating efficacy in various task, DeepSeq architecture based on a recursive Directed Acyclic Graph Neural Network (RecDAG-GNN~\cite{amizadeh2018learning}) results in extensive execution times, limiting its applicability for large-scale sequential designs. Furthermore, 
 the fixed number of iterations in DeepSeq's processing fails to adequately capture temporal correlations, particularly when the number of cycles in a simulated workload exceeds these iterations. This limitation is exacerbated by the model’s inability to differentiate effectively between different sequential behaviors and functionalities when nodes share similar logic and transition probabilities. 

More importantly, the use of a common embedding space for representing structural connectivity, logic function computation, and sequential behavior restricts DeepSeq’s capacity to distinguish among these diverse aspects. Recent works demonstrate the benefits of \textit{Disentangled Representation Learning (DRL)} in various domains, that aims to disentangle and identify hidden factors in data for creating more nuanced representations~\cite{wang2024disentangledrepresentationlearning,hsieh2018learning}. Motivated by this, we argue that an effective sequential circuit representation requires a model that can not only integrate but also clearly differentiate the above mentioned components based on their distinct properties. 

To address the above limitations, we introduce DeepSeq2, a robust framework engineered for efficient and effective representation learning of sequential netlists. Recognizing that the structure, function, and sequential behavior of circuits, while interrelated, involve distinct computational processes, DeepSeq2 distinctively learns and maintains these components in separate embedding spaces. This separation facilitates a more comprehensive representation, enabling the novel Directed Acyclic Graph Neural Network (DAG-GNN) architecture of DeepSeq2 to retain inter-dependencies while efficiently processing each aspect of the circuit's information.
DeepSeq2 also introduces a directed propagation scheme, specifically crafted to handle the challenges posed by cyclic sequential netlists. This scheme ensures the temporal correlations between components are captured effectively in a single forward pass, substantially reducing the training and inference times for large-scale circuit designs, thereby addressing the computational inefficiencies of its predecessor.
Moreover, DeepSeq2 employs a comprehensive set of supervisory tasks to accurately learn across different facets of the circuit. Drawing from the methodologies established by DeepGate2~\cite{shi2023deepgate2}, it utilizes reconvergence fanouts, logic-1 probability, and pairwise truth-table differences as node-level supervisions for structure and function learning. These supervisions enrich the model's understanding of complex circuit structures and functions, improving the precision of the learned embeddings. 
Additionally, to accurately capture the sequential behavior, DeepSeq2 employs a novel supervisory task: pairwise flip-flop (FF) similarity, that assesses the occurrence of identical state transitions for FF pairs under the same input conditions, enabling the model to differentiate between elements that exhibit similar transition probabilities but distinct sequential behaviors.
DeepSeq2 outperforms DeepSeq by showing better accuracy and runtime efficiency for complex sequential learning tasks such as power estimation and reliability analysis.
\vspace{-3pt}\section{Related Work}\label{sec:related}
\begin{figure}[!t]
    \vspace{-10pt}
	\centering
 \includegraphics[width=0.8\linewidth]{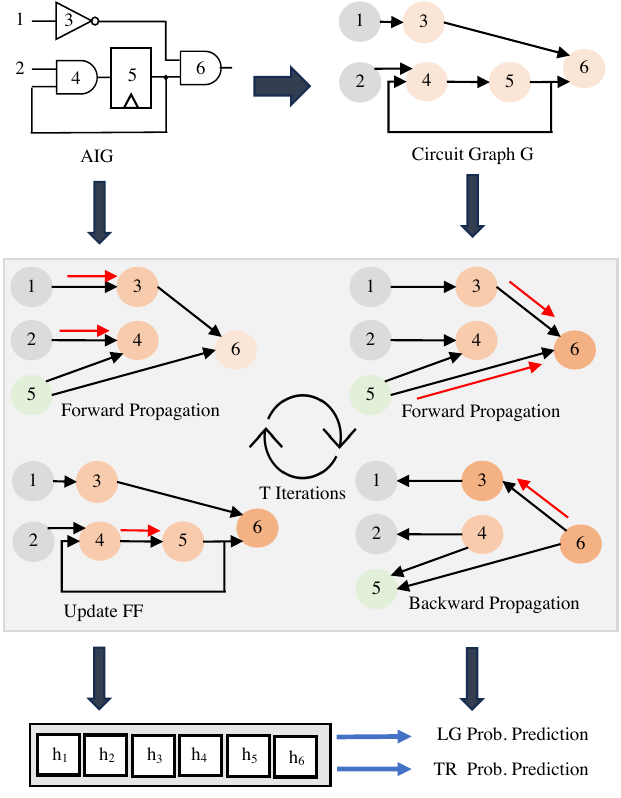}
	\caption{The overview of DeepSeq framework}
	\label{FIG:deepseq}
  \vspace{-15pt}
\end{figure}
\subsection{Circuit Representation Learning}
\begin{figure*}[!t]
    \vspace{-10pt}
	\centering
\includegraphics[width=0.95\linewidth]{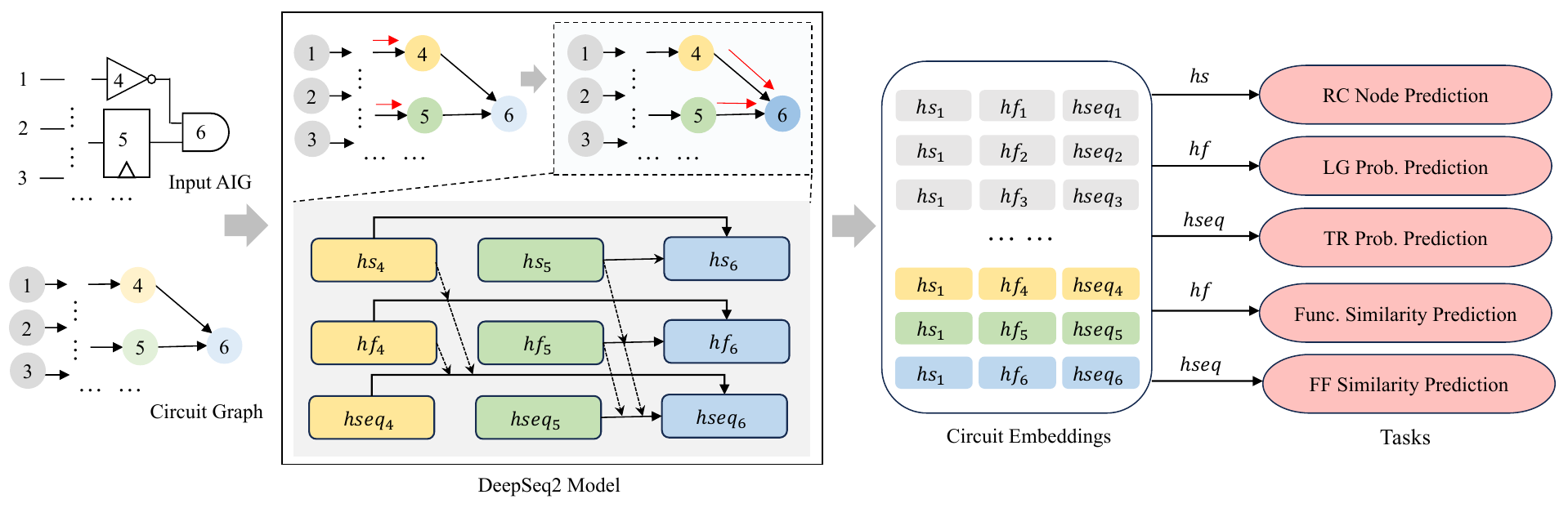}
	\caption{The overview of DeepSeq2 framework}
	\label{FIG:arch}
	\vspace{-7pt}
\end{figure*}
GNNs have greatly advanced gate-level combinational netlist representation learning, leveraging their suitability for modeling the graph-like structure of circuits
\cite{welling2016semi, hamilton2017inductive,zhang2019dvae,thost2021directed,amizadeh2018learning}. For example, ABGNN \cite{he2021graph} effectively encodes the structural aspects of netlists, enhancing tasks like arithmetic block identification. 
DeepGate~\cite{li2022deepgate} utilizes attention-based aggregation function and skip connections to capture the logic computations within combinational  and improve solutions for test point insertion\cite{shi2022deeptpi} and boolean satisfiability~\cite{li2023eda}.
FGNN \cite{wang2022functionality} employs contrastive learning to represent netlists at the graph level, facilitating netlist classification and subnet identification. Whereas, Gamora \cite{wu2023gamora}, utilizes high-level functional block reasoning to extend the capabilities of GNNs within the combinational circuit domain. DeepGate2~\cite{shi2023deepgate2} enhances DeepGate~\cite{li2022deepgate} with pairwise truth table differences of logic gates as supervision, and a functionality-aware loss function, along with a streamlined one-round DAG-GNN architecture to improve processing speeds and scalability for larger combinational circuit designs. HOGA~\cite{deng2024less} utilizes self-attention to aggregate node features, demonstrating its adaptability to QoR prediction and functional reasoning tasks. The advancements in the field have primarily focused on combinational netlists. DeepSeq~\cite{khan2023deepseq} is the sole framework tailored for learning sequential circuit representation as described below:
\vspace{-6.5pt}
\subsection{DeepSeq Framework}
DeepSeq is a GNN-based representation learning framework for sequential netlists. It captures temporal correlations between gates and flip-flops using a customized propagation scheme along with a multi-task training objective with logic-1 probabilities and state transition probabilities as supervisions. It also introduces the \emph{Dual Attention} aggregation function to optimize the learning. While it shows promise in power estimation and reliability analysis, DeepSeq has notable limitations.
Three critical aspects must be considered for high-quality sequential circuit representations:
\vspace{-2pt}
\begin{enumerate}
    \item \textbf{Logic Function Computation}: Depends on current input and state of netlist, leads to accurate behavior prediction.
    \item \textbf{State Transitioning Behavior}: Reflects the temporal correlations dictated by state changes in the netlist across consecutive clock cycles, crucial for dynamic analysis.
    \item \textbf{Circuit Structure}: Provides static information about the connectivity among circuit elements, essential for learning spatial correlations within the sequential netlist.
\end{enumerate}
\vspace{-2pt}
Due to their distinct computational process, a well-rounded sequential circuit representation model should differentiate these aspects while preserving their inherent interconnections. DeepSeq, however, consolidates all aspects into a shared embedding space, which considerably restricts its ability to distinguish among them.
Furthermore, DeepSeq's architecture is grounded in a recursive Directed Acyclic Graph Neural Network (RecDAG-GNN) \cite{amizadeh2018learning}, designed to manage temporal correlations within cyclic sequential netlists. This approach introduces certain limitations:
\vspace{-3pt}
\begin{itemize}
    \item The recursive nature of DeepSeq significantly increases runtime, making it less feasible for large-scale designs.
    \item Fixed number of $10$ propagation iterations in DeepSeq limits its ability to capture circuit behaviors beyond $10$ clock cycles, potentially missing longer temporal dynamics.
\end{itemize}
\vspace{-3pt}
Finally, DeepSeq's supervisory mechanism, based on logic-1 probability and transition probabilities, fails to adequately account for the variances in functions and sequential behaviors of nodes with identical probabilities, highlighting a need for improvement.
\vspace{-4pt}

\section{METHODOLOGY}\label{sec:method}
This section presents DeepSeq2 framework, which aims to enhance the representation learning of sequential netlists by addressing the distinct yet interconnected components of circuit information: structure, function, and sequential behavior.
\vspace{-8pt}
\subsection{Problem Formulation}

DeepSeq2 strategically divides the representation of sequential circuit information into three specialized embedding spaces (Figure~\ref{FIG:arch}) to isolate and enhance the learning of each aspect:
\vspace{-2pt}
\begin{itemize}
    \item \textbf{Structure Information}: Captures the intricate interconnections between circuit elements, crucial for understanding the architectural dependencies within the circuit.
    \item \textbf{Function Information}: Represents the logical computations performed by these elements, which depend on both the input signals and the current state of the circuit.
    \item \textbf{Sequential Behavior}: Characterizes the nuanced patterns of state transitions between consecutive clock cycles, essential for capturing the dynamic nature of the circuits.
\end{itemize}
\vspace{-2pt}
By segregating these components into disentangled embeddings, DeepSeq2 ensures a nuanced representation, preserving the unique characteristics of each while maintaining their interdependencies.
\vspace{-7pt}
\subsection{Training Objective}
To encode the distinct aspects of the netlist effectively, we utilize a multi-task training strategy that enhances the specificity and accuracy of the learned models.

\noindent\textbf{Structure Encoding:}
Understanding reconverging fanouts (where multiple outputs merge) is crucial for grasping circuit complexity as they reveal structural dependencies. DeepSeq2, inspired by DeepGate2 \cite{shi2023deepgate2}, predicts gate reconvergence ($\mathcal{T}^{RC}$) to identify critical circuit junctures, enhancing structural representation.

\noindent\textbf{Function Encoding:}
The function encoding of a sequential netlist is determined by the combinational components, which are influenced by both primary inputs (PIs) and pseudo primary inputs (PPIs), emanating from the FFs. DeepSeq2 supervises each node with logic-1 probabilities under random inputs, represented as $\mathcal{T}^{LG}$ to learn the function information~\cite{li2022deepgate, shi2023deepgate2, khan2023deepseq}.
While logic-1 probabilities represents the functional behavior of a netlist, it cannot differentiate between nodes that, despite sharing the same probability values, perform different functions. This limitation is significant in complex circuits where distinct functional pathways may have similar probabilistic behavior but differ in their functionality. 

To overcome this challenge, similar to the enhancements introduced in DeepGate2 \cite{shi2023deepgate2}, DeepSeq2 incorporates pairwise truth table differences as an additional layer of node-level supervision, ($\mathcal{T}^{F}$). This method involves comparing the truth tables of pairs of logic gates to discern subtle differences in their functionalities, thereby enabling the model to learn and encode nuanced differences in similar-looking functions. This supervision is crucial for teaching the model to recognize and differentiate between functionally distinct but probabilistically similar logic operations.
DeepSeq2 applies pairwise truth table difference supervision to combinational logic components only and logic-1 probability supervision to all gates and flip-flops in the netlist. 
This comprehensive application is critical as the netlist's sequential behavior relies on accurate logic function computation over successive cycles.

\noindent\textbf{Sequential Encoding:}
To capture the sequential behavior of a netlist, accounting for temporal correlations between gates and flip-flops is essential. Transition probabilities, used in DeepSeq~\cite{khan2023deepseq}, provide a foundational supervisory signal, but they may not fully represent the complexity of state transitions in sequential circuits. For example, FFs with identical transition probabilities can exhibit significantly different state transitioning behaviors over time.

To address these nuances, we have developed a novel supervisory approach that not only includes transition probabilities prediction \textbf{($\mathcal{T}^{TR}$)} but also introduces a unique metric for encoding the pairwise state-transition similarity of FFs, termed \textbf{$\mathcal{T}^{FF_{sim}}$}. This approach defines two indicators $\phi_{t}^{state(i, j)}$ and $\phi_{t}^{trans(i, j)}$ for each pair of FFs $(i, j)$ at every clock cycle $t$.  $\phi_{t}^{state(i, j)}$ will be $1$ if FFs $i$ and $j$ have the same state and $0$ otherwise. Similarly, $\phi_{t}^{trans(i, j)}$ will be $1$ if FFs $i$ and $j$ not only have the same state but also transition to same state and $0$ otherwise.
The state-transition similarity for FFs $(i, j)$ can be quantified using the ratio of matching state transitions to same states across all evaluated cycles, as shown below:
\vspace{-5pt}
\begin{equation}
sim_{i, j} = \frac{\sum_{t \in T} \phi_{t}^{trans(i, j)}}{\sum_{t \in T} \phi_{t}^{state(i, j)}}
\vspace{-7pt}
\end{equation}
 $\mathcal{T}^{FF_{sim}}$ encourages the model to represent similar sequential behaviors closely in the embedding space. Notably, \textbf{$\mathcal{T}^{FF_{sim}}$} is applied on pairs of FFs while \textbf{$\mathcal{T}^{TR}$} is applied to all gates and FFs for a comprehensive understanding of interdependencies between logic function computation and state transitioning dynamics in circuit.

\vspace{-7pt}
\subsection{GNN model}
\begin{figure*}[!t]
    \vspace{-10pt}
	\centering
 \includegraphics[width=1.0\linewidth]{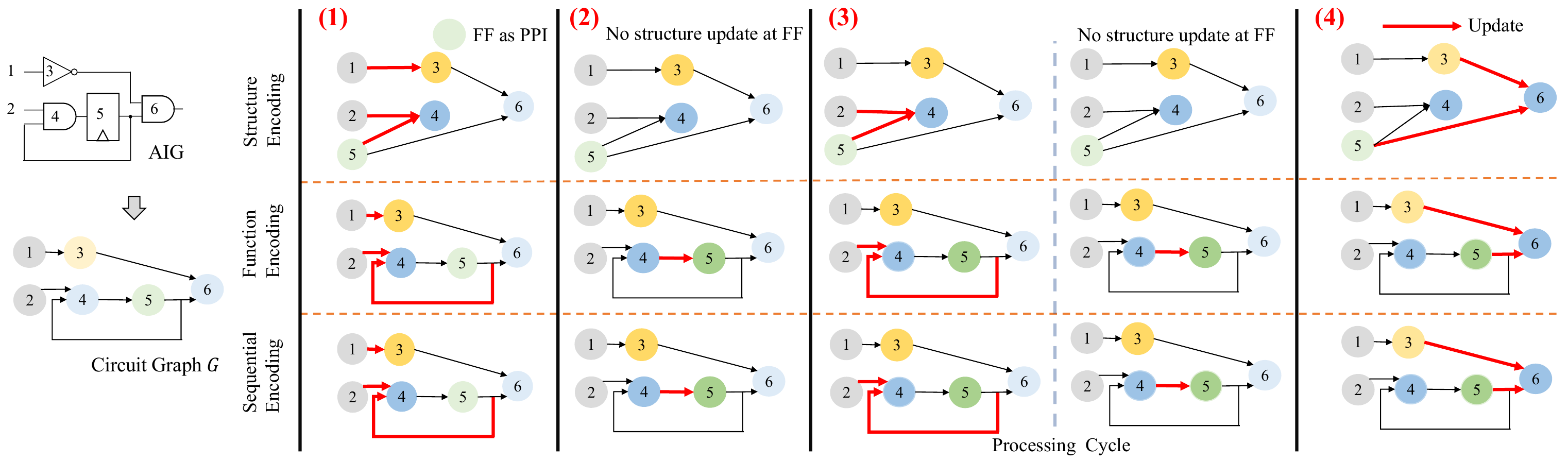}
	\caption{The overview of  propagation scheme.}
	\label{FIG:prop}
	\vspace{-5pt}
\end{figure*}


The core learning framework of DeepSeq2 is based on DAG-GNN~\cite{thost2021directed, zhang2019dvae}, specifically engineered to handle cyclic sequential netlists efficiently in a single forward pass, addressing runtime inefficiencies observed in its predecessor, DeepSeq~\cite{khan2023deepseq}. The circuit is represented as a directed cyclic graph $G$ where $V$ represents all nodes $v$ in the graph to accurately reflect the circuit's topology and operational dependencies.
Embeddings are intialized as follows:
\vspace{-2pt}
\begin{itemize}
    \item \textbf{Structure Embeddings ($h_s$)}: For PIs and flip-flops (considered as PPIs), structure embeddings are initialized with orthogonal vectors as per DeepGate2 \cite{shi2023deepgate2} for effective structural learning. Other nodes receive random initial vectors.
    \item \textbf{Function and Sequential Embeddings ($h_f$, $h_{seq}$)}: For PIs, these embeddings are initialized based on the logic-1 probabilities and transition probabilities corresponding to the predefined workload. This approach ensures that the initial state reflects real operational conditions. For non-PI nodes, $h_f$ and $h_{seq}$ are initialized randomly.
\end{itemize}
\vspace{-2pt}
During training, DeepSeq2 keeps $h_s$ constant for PIs and PPIs, updates $h_s$ for other nodes to refine circuit structure understanding, and dynamically adjusts $h_f$ and $h_{seq}$ for all nodes except PIs to enhance inference learning under specified workloads.
DeepSeq2 uses a tailored propagation strategy to efficiently navigate cyclic structures in sequential netlists, enabling accurate learning of temporal correlations in a single pass as follows (see Figure~\ref{FIG:prop}):

\begin{enumerate}
    \item Begin at PIs. Progress sequentially through the directed graph $G$, updating the $h_s$, $h_f$, and $h_{seq}$ for each logic gate.
    \item Upon reaching FFs, update only their $h_f$ and $h_{seq}$. $h_s$ for FFs remain unchanged as they are considered PPIs and are critical for maintaining consistent structural learning.
    \item Detect and address cycles formed by newly updated FFs. If cycles are identified, isolate and process the implicated sub-circuit iteratively using steps (1) and (2), which is crucial for capturing the cyclic dependencies accurately.
    \item Continue this propagation through all graph levels until reaching primary outputs.
\end{enumerate}
During propagation, $aggregate$ and $combine$ operations, differentiated based on node types to accommodate varying logic behaviors, are executed to refine node embeddings. 
\vspace{-3pt}
\begin{itemize}
    \item \textbf{Aggregate}: Collects and synthesizes information from preceding nodes, adapting to the specific requirements of structure, function, or sequential data.
    \item \textbf{Combine}: Integrates aggregated data to update the node’s state, reflecting the unique characteristics of each node type.
\end{itemize}

During this process, we calculate the structure embeddings of all nodes $v$ except PIs and PPIs as follows, where $x \in \{AND , NOT\}$.
\vspace{-2pt}
\begin{equation} \label{Eq:hs}
    \begin{split}
        hs_{v} & = aggr_{x}^{s} (hs_j | j \in \mathcal{P}(v)) \\
    \end{split}
\end{equation}
To aggregate the function-related information for a node $v$, we concatenate the structure and function embeddings of its predecessors, considering the crucial impact of node's connectivity on its function as below, where $x \in \{AND , NOT, FF\}$:
\vspace{-2pt}
\begin{equation} \label{Eq:hf}
    \begin{split}
        hf_{v} & = aggr_{x}^{f} ([hs_j, hf_j] | j \in \mathcal{P}(v)) \\ 
    \end{split}
    \vspace{-5pt}
\end{equation}
The sequential behavior captures the circuit output at two consecutive clock cycles, which depends upon the circuit connectivity and logic computation process. Therefore, to encode the sequential behavior of node $v$, we aggregate the structure, function, and sequential embeddings of its predecessors as follows:
\vspace{-2pt}
\begin{equation} \label{Eq:hf}
    \begin{split}
        h{seq}_{v} & = aggr_{x}^{f} ([hs_j, hf_j, h{seq}_j] | j \in \mathcal{P}(v)) \\ 
    \end{split}
\end{equation}
Each aggregate function in the above equations is implemented using attention mechanism~\cite{velikovi2017graph,zhang2020graph, li2022deepgate,shi2023deepgate2,khan2023deepseq}.
\begin{equation}
\small
    \mathbf{m}_{v} = \sum_{u \in \mathcal{P}(v)}  \alpha_{uv}^t  \mathbf{h}_u^t
\text{\quad where\quad} \alpha_{uv}^t = \mathop{softmax}\limits_{u \in \mathcal{P}(v)} (w_1^\top \mathbf{h}_v^{t-1}+ w_2^\top \mathbf{h}_u^t)
\label{eq:attn1}
\end{equation}
where $\alpha_{uv}^t$ is a weighting coefficient that learns the impact of predecessors' information on a node $v$, since the predecessors with controlling values influence the output of logic gate more than the predecessors with non-controlling values. 
After aggregating the information, we use gated recurrent unit (GRU)~\cite{li2022deepgate,zhang2019dvae,thost2021directed} to update the hidden state of a node $v$. Similar to aggregate function, we instantiate separate GRU function for each node type.

\begin{table}
\centering
\scriptsize
\captionsetup{font=small}
\caption{The Statistics of Training Dataset} \label{TAB:TrainData}
\vspace{-10pt}
\setlength\tabcolsep{15.0pt}
\begin{tabular}{@{}l|l|l@{}}
\toprule
Benchmark & \# Subcircuits & \# Nodes (Avg. $\pm$ Std.) \\ \midrule
ISCAS'89~\cite{ISCAS89}  & $1,159$           & $148.88 \pm 87.56$        \\
ITC'99~\cite{ITC99}    & $1,691$           & $272.6 \pm 108.33$        \\
Opencores~\cite{takeda2008opencore} & $7,684$           & $211.41 \pm 81.37$        \\ \midrule
\textbf{Total} & $10,534$ & $214.35 \pm 92.63$ \\
\bottomrule
\end{tabular}
\vspace{-10pt}
\end{table}
\begin{table}
\centering
\scriptsize
\captionsetup{font=small}
\caption{Performance Comparison b/w DeepSeq and Baselines}
\vspace{-10pt}
\setlength\tabcolsep{15.0pt}
\label{TAB:Comparison}
\begin{tabular}{@{}c|c|c|c@{}}
\toprule
\multirow{2}{*}{Model}       & \multirow{2}{*}{Aggregation} & Avg. PE & Avg. PE \\
                             &                              & ($\mathcal{T}^{TR}$) & ($\mathcal{T}^{LG}$)       \\ \midrule
\multirow{2}{*}{DAG-ConvGNN} & Conv. Sum                    & 0.066   & 0.236   \\
                             & Attention                    & 0.065   & 0.220   \\ \midrule
\multirow{2}{*}{DAG-RecGNN}  & Conv. Sum                    & 0.045   & 0.104   \\
                             & Attention                    & 0.035   & 0.095   \\ \midrule
DeepSeq                      & Dual Attention               & 0.028  & 0.080  \\ \midrule
 DeepSeq2 & Attention & \textit{$0.021$}           & \textit{$0.036$} \\
\bottomrule
\end{tabular}
\vspace{-13pt}
\end{table}
\section{Experiments}\label{sec:exp}

\subsection{Experimental Settings}
\label{Sec:Exp:Setting}
\noindent \textbf{Dataset:}\label{Sec:Exp:Dataset}
We use the dataset from DeepSeq~\cite{khan2023deepseq}, consisting of $10,534$ sequential sub-circuits extracted from open-sourced benchmarks: ISCAS'89~\cite{ISCAS89}, ITC'99~\cite{ITC99}, and Opencore~\cite{takeda2008opencore} (Table~\ref{TAB:TrainData}).
 The circuits are transformed into and-inverter(AIG) graphs and optimized using ABC\cite{brayton2010abc} for a uniform design distribution for learning process of the model, as demonstrated in~\cite{li2022deepgate,shi2023deepgate2,khan2023deepseq}.The resulting circuit consists of $4$ types of nodes: primary inputs, AND gates, NOT gates, and flip-flops. 
To generate the supervision for $\mathcal{T}^{RC}$, we analyze the two fan-in gates in the circuit. If the fan-ins of a gate reconverge to a predecessor gate, we assign a binary label of $1$ to the fan-in gate pair. Otherwise, the gate pair is labeled as $0$. 
For the $\mathcal{T}^{LG}$ and $\mathcal{T}^{TR}$, we simulate $1,000$ random sequential patterns, each consists of $100$ cycles and calculate the transition probability and logic probability for each gate.
 For $\mathcal{T}^{F}$,  we follow the sampling conditions from DeepGate2~\cite{shi2023deepgate2} and sample $5.72$M  pairs of logic gates with an average of $543.17$ pairs per circuit to calculate pair-wise truth-table differences. 
 Lastly, we select $5.22$M pairs of FFs, with an average of $495.26$ pairs per circuit, for predicting \textbf{$\mathcal{T}^{FF_{sim}}$} similarity. 
 To ensure the quality of FF pairs, we only consider the FF pairs that are driven by the same primary inputs and subjected to the same clock delay. 
\begin{table}
\centering
\captionsetup{font=small}
\caption{Effectiveness of Different Components of DeepSeq2}
\vspace{-8pt}
\label{TAB:Ablation}
\scriptsize
\renewcommand\tabcolsep{10pt}
\begin{tabular}{@{}l|l|l|l|l@{}}
\toprule

\multirow{2}{*}{Model}&  \multirow{2}{*}{Embedding} &  \multirow{2}{*}{Propagation}  &  Avg. PE & Avg. PE \\
                    &&    & ($\mathcal{T}^{TR}$) & ($\mathcal{T}^{LG}$)       \\ \midrule
                        
DeepSeq2  & Combine & Custom (DeepSeq~\cite{khan2023deepseq}) & 0.026  & 0.070
                             \\ 
DeepSeq2  & Combine & Our & 0.023 & 0.048   \\
DeepSeq2  & Separate & Our & 0.021 & 0.036   \\ 
 \bottomrule
\end{tabular}
\vspace{-10pt}
\end{table}
\begin{table}[!t]
\centering
\captionsetup{font=small}
\caption{Power Estimation on 6 Large-Scale Circuits} \label{TAB:TestPE}
\vspace{-10pt}
\scriptsize
\renewcommand\tabcolsep{3.5pt}
\begin{tabular}{@{}ll|l|lll|lll@{}}
\toprule
\multirow{2}{*}{Design  Name}              & \multirow{2}{*}{\# Nodes}    & GT     & DeepSeq   & \multirow{1}{*}{Error} & Time & DeepSeq2  & \multirow{1}{*}{Error}&   Time 
 \\ 
 &                  & (mW) & (mW)   &  & (sec) & (mW)  & &   (sec) 
 \\ \midrule
noc\_router     & 5,246    &{0.653}                 & 0.643       & 1.53\% & 14.12  & 0.652 &   0.153\% &  6.28 \\
pll              & 18,208        & {0.936}                 & 0.960        & 2.56\%    &    107.58 & 0.924 & 1.28\% & 16.11\\
ptc      & 2,024    & {0.247}              & 0.239        & 3.24\%       & 7.34 & 0.242 & 2.02\%  &1.05\\
rtcclock      & 4,720     & {0.463}                     & 0.442        & 4.54\%  &  24.52     & 0.456 & 1.51\% &3.58\\
ac97\_ctrl    & 14,004  & {3.353}                 & 3.261        & 2.74\%   &  87.33    & 3.318 &  1.04\% &13.23 \\
mem\_ctrl            & 10,733  & {1.365}                  & 1.303        & 4.54\%       & 69.01 & 1.345 & 1.46\%  &9.25\\ \midrule
\textbf{Avg.}                              &                                       & &&\textbf{3.20\%} &\textbf{51.65}&&\textbf{1.24\%}&\textbf{8.25}\\ \bottomrule
\end{tabular}
\vspace{-13pt}
\end{table}
\vspace{2pt}

\noindent\textbf{Implementation Details and Evaluation Metric:}
\label{Sec:params}
We compare DeepSeq2 with DAG-ConvGNN~\cite{zhang2019dvae,thost2021directed}, DAG-RecGNN~\cite{amizadeh2018learning} and DeepSeq.
Each model has one forward and one reverse layer. For DAG-ConvGNN and DAG-RecGNN, we try two different aggregation functions, i.e., convo-
lutional sum (conv. sum)~\cite{welling2016semi}, attention~\cite{thost2021directed}. For DAG-RecGNN and DeepSeq, we use 10 iterations to
obtain final embeddings.
We implement all models using pytorch geometric (PyG)~\cite{pyg} framework. The dimensions for $\mathbf{h}_s$, $\mathbf{h}_f$ and $\mathbf{h}_{seq}$  are 128. We also have 3 regressors, each consists of 3-MLPs for the prediction of $\mathcal{T}^{RC}$, $\mathcal{T}^{LG}$ and $\mathcal{T}^{TR}$ tasks. We use Rectified linear unit (ReLU) function as an activation function between MLP layers. DeepSeq2 is trained using the ADAM optimizer and $16$ batch size for $80$ epochs with a $1\times10^{-4}$ learning rate. We also incorporated the topological batching method from~\cite{thost2021directed} to speed up the training as in~\cite{li2022deepgate, shi2023deepgate2, khan2023deepseq}.
We use BCE loss for \textbf{$\mathcal{T}^{RC}$},   
 L1 losses for \textbf{$\mathcal{T}^{LG}$} and \textbf{$\mathcal{T}^{TR}$} and  functionality-aware loss from DeepGate2~\cite{shi2023deepgate2} for \textbf{$\mathcal{T}^{F}$} and \textbf{$\mathcal{T}^{FF_{sim}}$}. The final loss is the weighted sum of individual loss functions.
DeepGate2~\cite{shi2023deepgate2} already encodes the combinational netlist into separate structure and function embeddings using  \textbf{$\mathcal{T}^{RC}$}, \textbf{$\mathcal{T}^{LG}$},\textbf{$\mathcal{T}^{F}$}, we benefits DeepSeq2 from it by leveraging it as a pre-trained model.
Specifically, we initialize the weight of all aggregate and update operations designated for learning structure and function embeddings from the weights of their respective operations from DeepGate2. 
Then, along with \textbf{$\mathcal{T}^{RC}$}, \textbf{$\mathcal{T}^{LG}$},\textbf{$\mathcal{T}^{F}$}, we add one more \textbf{$\mathcal{T}^{TR}$} task and train it for $40$ epochs.
After that,
we add \textbf{$\mathcal{T}^{FF_{sim}}$} in the training process for next $40$ epochs. 
This is inspired by curriculum learning process. We consider \textbf{$\mathcal{T}^{FF_{sim}}$} as the hardest task among all add it later in the learning process. 
To compare the results with baseline models, we use \textit{average prediction error} (Avg. PE), which is defined as the average value of the absolute differences between the ground-truth ($y_v$) and the predictions ($\hat{y}_v$) from the model.
\vspace{-6pt}
\subsection{Results}
\label{Section:Exp:Results}
\noindent\textbf{Performance Comparison b/w DeepSeq and Baselines:}
The results in Table~\ref{TAB:Comparison} demonstrates the superior performance of DeepSeq compared to DAG-GNN and DAG-RecGNN models in terms of avg. PE for both $\mathcal{T}^{LG}$ and $\mathcal{T}^{TR}$. However, with our dedicated propagation scheme and informative set of supervisions, DeepSeq2 outperforms the DeepSeq. For both $\mathcal{T}^{LG}$ and $\mathcal{T}^{TR}$, it achieves $25.00\%$ and $55.00\%$ relative improvement on avg. PE, respectively.
The significant performance improvement for $\mathcal{T}^{LG}$ shows that DeepSeq2 can effectively handle the correlations between combinational and sequential components in the netlist. Losses of remaining tasks, i.e., $\mathcal{T}^{RC}$, $\mathcal{T}^{F}$ and $\mathcal{T}^{FF_{sim}}$, are $0.20$, $0.08$ and $0.071$, respectively. We do not add it in Table~\ref{TAB:Comparison} since, DeepSeq does not support these tasks.

\vspace{2pt}
\noindent\textbf{Effectiveness of different components of DeepSeq2:}
Table~\ref{TAB:Ablation} demonstrates the effectiveness of different components in DeepSeq2 compared to DeepSeq. In the first setting, we modify the supervision while keeping the customized propagation from DeepSeq. Here, DeepSeq2 achieves improved performance solely by changing the supervision, surpassing DeepSeq. The average PE error for $\mathcal{T}^{LG}$ and $\mathcal{T}^{TR}$ is reduced by $7.14\%$ and $12.50\%$ respectively. In the next setting, we replace the customized propagation from DeepSeq with our proposed propagation mechanism. 
This further improves the performance by $11.54\%$ and $31.43\%$ for $\mathcal{T}^{LG}$ and $\mathcal{T}^{TR}$ respectively.
In our final experiment setting, we replace the combined embedding for structure, function, and sequential information with disentangled embeddings for each of them. his experiment brings the best result proving the effectiveness of our proposed method.

\begin{table}
\centering
\captionsetup{font=small}
\caption{Power Estimation on ac97\_ctrl with Different Workloads} 
\vspace{-5pt}
\label{TAB:PEac97}
\scriptsize
\renewcommand\tabcolsep{12pt}
\begin{tabular}{@{}ll|ll|ll@{}}
\toprule
Workload ID   & {GT }                 & DeepSeq & Error.  & DeepSeq2 & Error.          \\ \midrule
W0            & {3.353}                  & 3.261     & 2.74\%  &3.313    & 1.19\%   \\
W1            & {3.349}                  & 3.219     & 3.88\%  & 3.281  & 2.03\%      \\
W2            & {2.758}                & 2.697     & 2.21\%   & 2.723 &1.26\%      \\
W3            & {3.414}               & 3.322     & 2.69\%  &3.365 & 1.44\%        \\
W4            & {6.696}                   & 6.607     & 1.33\%   & 6.636  & 0.90\%     \\ \midrule
\textbf{Avg.} & \textbf{}                  &          \textbf{} & \textbf{2.57\%} && \textbf{1.36\%}\\ \bottomrule
\end{tabular}
\vspace{-17pt}
\end{table}

\vspace{-3pt}\section{Downstream Tasks}\label{sec:tasks}
This section evaluates the generalizability and accuracy of our pre-trained DeepSeq2 model on two downstream tasks: power estimation and reliability analysis. For both tasks, we test DeepSeq2 using the same circuits as in DeepSeq~\cite{khan2023deepseq} (see Table~\ref{TAB:TestPE}), which are 1-2 order of magnitude larger than the circuits used for pre-training.

 \vspace{-6pt}
\subsection{Evaluation on Power Estimation}
We evaluate DeepSeq2 on netlist-level \textit{dynamic power estimation}~\cite{tsui1995power} task that is calculated as $P = \frac{1}{2} C V_{dd}^2 y_{avg}^{TR}$ where $C$ is the capacitance, $V_{dd}$ is the supply voltage, and $y_{avg}^{TR}$ is the average transition probability of all gates. Recent DL-based power estimation solutions include GNN-based approaches
for FPGA HLS~\cite{lin2022powergear}, CNN-based methods for estimating power
for ASIC designs~\cite{zhou2019primal}.
The difference between our work and these are that they are specifically designed for power estimation method whereas DeepSeq2 is a generic representation learning framework, eligible to solve multiple downstream tasks related to circuit structure and functionality.
To our knowledge, Grannite~\cite{zhang2020grannite} is the only DAG-GNN based model for netlist-level power estimation. However, due to the superior performance demonstrated by DeepSeq~\cite{khan2023deepseq} over Grannite, we choose DeepSeq as our preferred baseline model for power estimation.
\begin{table}[]
\centering
\captionsetup{font=small}
\caption{The Results of Reliability Analysis} 
\label{TAB:Relia}
\vspace{-10pt}
\scriptsize
\renewcommand\tabcolsep{10pt}
\begin{tabular}{@{}ll|ll|ll@{}}
\toprule
Design Name   & GT          & DeepSeq & Error.  & DeepSeq2 & Error        \\ \midrule
noc\_router   & 0.9876        & 0.9814  & 0.63\%   &0.9834 & 0.43\%        \\
pll           & 0.9792         & 0.9857  & 0.66\%    & 0.9781 &0.11\%      \\
ptc           & 0.9970         & 0.9928  & 0.42\%   & 0.9951 & 0.19\%       \\
rtcclock      & 0.9985         & 0.9969  & 0.16\%  & 0.9974 & 0.11\%        \\
ac97\_ctrl    & 0.9953           & 0.9943  & 0.10\%  & 0.9946& 0.07\%        \\
mem\_ctrl     & 0.9958          & 0.9936  & 0.22\% &0.9947& 0.11\%        \\ \midrule
\textbf{Avg.} &       &         & \textbf{0.37\%} &  & \textbf{0.17\%} \\ \bottomrule
\end{tabular}
\vspace{-22pt}
\end{table}
\vspace{2pt}
\noindent\textbf{Experimental Settings:}
To maintain consistency with DeepSeq, we replicate its settings and fine-tune all parameters of DeepSeq2 on large test circuits using $100$ random workloads. This aligns with DeepSeq's analysis, which highlights distinctions in transition probability distributions between large test circuit designs and the small circuits in training dataset, due to low power design considerations~\cite{kathuria2011review}.
After fine-tuning, DeepSeq2 demonstrates the ability to generalize to an arbitrary number of unseen workloads for that circuit. To accommodate our model's AIG format input requirement, we transform all gate types in our test circuits (Table~\ref{TAB:TestPE}) into combinations of AND and NOT gates. Consequently, we only record the switching activities of fanout gates in the resulting combinations for power estimation, as these gates retain the same switching activity as the original ones. During inference, we collect logic and transition probabilities for each primary input (PI) from the testbench files of the corresponding test circuit. Based on that we generate the workload for the underlying circuit (see Section \ref{Sec:Exp:Dataset} for more details).
Ground truth (GT), DeepSeq (baseline), and DeepSeq2 transition probabilities are converted into SAIF files, which are provided to a commercial power analysis tool utilizing a TSMC 90nm standard cell library. This process yields average power consumption values: GT Power, DeepSeq Power, and DeepSeq2 Power.

\vspace{1pt}
\noindent\textbf{Power Estimation on the large-scale circuits:}
Table~\ref{TAB:TestPE} shows the accuracy and runtime improvement achieved by DeepSeq2 over DeepSeq, where the estimated power based on transition probabilities computed from ground-truth, DeepSeq, and DeepSeq2 are abbreviated as GT, DeepSeq, and DeepSeq2, respectively. Overall, the  DeepSeq produces $3.20\%$ whereas DeepSeq2 only shows $1.24\%$ error on average, which is much closer to the ground-truth power estimation. To conclude, DeepSeq2 brings a significant improvement in error reduction, i.e., $61.25\%$ over DeepSeq, on large-scale circuits. Besides this, DeepSeq2 also shows significant runtime improvement over DeepSeq. Consider the largest circuit in our test data, i.e., \textit{pll}. DeepSeq takes $107.58$ seconds to infer it whereas DeepSeq2 only takes $16.11$ seconds. 
This improvement is attributed to the one time processing mechanism of sequential netlist in DeepSeq2, as opposed to DeepSeq's iterative approach. However, the gain in not an-order of magnitude due to the need for DeepSeq2 to re-process cyclic sub-netlists within the sequential netlist.


\vspace{2pt}
\noindent\textbf{Power Estimation under different workloads:}
Table~\ref{TAB:PEac97} illustrates DeepSeq2's superior generalizability over DeepSeq. Assessing power estimation for the \textit{ac97\_ctrl} circuit with 5 unseen workloads, DeepSeq2 achieves a remarkable error rate of $1.36\%$ compared to the ground truth, while DeepSeq yields a higher error rate of $2.57\%$.
\vspace{-13pt}
\subsection{Evaluation on Reliability Analysis}\label{sec:reliability}
In this section, we present the performance of our pre-trained DeepSeq2 model on netlist's reliability analysis task~\cite{mahdavi2009scrap}.

\vspace{2pt}
\noindent\textbf{Experimental Settings:}
To the best of our knowledge, DeepSeq~\cite{khan2023deepseq} is the only learning-based solution that targets reliability analysis problem.  Therefore, we select DeepSeq as the baseline for our comparative analysis in this experiment. 
To generate ground truth (GT) results, fault-free simulations and faulty simulations using a Monte Carlo method with a $0.05\%$ error rate involving $1000$ sequential patterns each spanning $100$ cycles are performed. 
By comparing the difference in responses from both simulations, the error probabilities are calculated.
Then, DeepSeq2 is fine-tuned in end to end manner for reliability analysis task by generating 2D node-level supervision for each circuit node using the dataset described in Table~\ref{TAB:TrainData}. This supervision represents the probabilities of bit flipping from $0$ to $1$ and from $1$ to $0$ and are generated in the same manner we generate GT results.
 . 
To capture the combined influence of a circuit's structure, function, and sequential information on fault propagation, we concatenate these embeddings for each node and replace the regressor in DeepSeq2 to predict the error probabilities.  We fine-tune DeepSeq2 for $50$ epochs using L1 loss, utilizing the same hyper-parameter configurations as described in Section~\ref{Sec:params}. 
\vspace{2pt}
\noindent\textbf{Results:}
Table~\ref{TAB:Relia} presents the superior performance of DeepSeq2 compared to DeepSeq where the reliability values computed from the ground-truth, DeepSeq, and DeepSeq2 are abbreviated as GT, DeepSeq, and DeepSeq2, respectively. 
DeepSeq achieves a low average error of $0.37\%$, compare to GT values. However, DeepSeq2 surpasses DeepSeq with a 54.05\% improvement by reaching an average error of just $0.17\%$.
This substantial improvement highlights the effectiveness of DeepSeq2 in capturing and leveraging the relevant information for more accurate and reliable estimations.
\vspace{-2pt}\section{Conclusion and Future Work}\label{sec:conclusion}

We present, DeepSeq2 a representation learning framework that encodes sequential netlists into structural, functional, and sequential embeddings. It utilizes an efficient DAG-GNN model to process cyclic sequential netlists in a single pass and incorporates pair-wise FF similarity supervision to accurately capture the state transitioning patterns. It outperforms DeepSeq~\cite{khan2023deepseq} in accuracy, runtime efficiency, and generalizability both in pre-training tasks like logic probability, transition probability prediction and downstream tasks such as power estimation, and reliability analysis.
Moreover, recent studies \cite{luo2024transformers, shirzad2023exphormer} highlight the effectiveness of applying transformers to graphs, enabling better discernment of complex relationships and scalability. Our future work aims to leverage transformers to enhance our model's ability to capture temporal correlations and gain a more nuanced understanding of time-dependent data patterns.
\vspace{-2pt}
\section{Acknowledgments}
This work was supported in part by the General Research Fund of the Hong Kong Research Grants Council (RGC) under Grant No. 14212422 and 14202824.

\clearpage
\balance
\bibliographystyle{unsrt}
\bibliography{main}

\end{document}